\documentclass[]{aa} 

\usepackage{graphicx}
\usepackage{txfonts}
\usepackage{natbib}
\usepackage{longtable}

\begin{document}

\title{
Profiles of interstellar cloud filaments.
}
\subtitle{Observational effects in synthetic sub-millimetre observations.}

\author{M.     Juvela\inst{1},
        J.     Malinen\inst{1},
        T.     Lunttila\inst{1}
}

\institute{
Department of Physics, P.O.Box 64, FI-00014, University of Helsinki,
Finland, {\em mika.juvela@helsinki.fi}
}

\authorrunning{M. Juvela et al.}


\abstract
{
Sub-millimetre observations suggest that the filaments of interstellar
clouds have rather uniform widths and can be described with the
so-called Plummer profiles. The shapes of the filament profiles are
believed to be closely linked to the formation mechanism and physical
state of the filaments.
}
{
Before drawing conclusions on the observed column density profiles, we
must evaluate the observational uncertainties. We want to estimate the
bias that could result from radiative transfer effects and from the
variations of submm dust emissivity that are expected to take
place in dense clouds.
}
{
We use cloud models obtained with magnetohydrodynamic simulations and 
carry out radiative transfer calculations to produce maps of
sub-millimetre surface brightness. Column densities are estimated
based on the synthetic observations. For selected filaments, the
estimated profiles are compared to those derived from the original
column density. Possible effects from spatial variations of dust
properties are examined.
}
{
With instrumental noise typical of the $Herschel$ observations, the 
parameters derived for nearby clouds are correct to within a few
percent. The radiative transfer effects have only a minor effect on
the results. If the signal-to-noise ratio is degraded by a factor of
four, the errors become significant and for half of the examined
filaments the values remain practically unconstrained. The errors also
increase in proportion to the cloud distance. Assuming the
resolution of Herschel instruments, the model filaments are barely
resolved at a distance of $\sim$400\,pc and the errors in the
parameters of the Plummer function are several tens of per cent. 
}
{
The Plummer parameters, in particular the power-law exponent $p$, are
sensitive to noise but can be determined with good accuracy using 
$Herschel$ data. However, one must be cautious about possible
line-of-sight confusion. In our model cloud, a large fraction of the
filaments identified from column density maps are not continuous
structures in three dimensions. 
}

\maketitle
%

\section{Introduction}

For a long time the filaments have been recognised as a common feature
of interstellar clouds that is also closely related to the star
formation process \citep[e.g.,][]{Elmegreen1979, Schneider1979,
Bally1987}. Recent \emph{Herschel} studies and ground based surveys have
revealed the ubiquity of filamentary structures within dense
interstellar clouds, and pre-stellar cores and protostars along dense, gravitationally unstable filaments
\citep[e.g.,][]{Menshchikov2010, Andre2010, Arzoumanian2011, PaperIII}. 
Therefore, the structure and formation of the filaments are important
parts of star formation studies. The filaments are naturally formed as
a result of interstellar turbulence
\citep[e.g.,][]{PadoanNordlund2011}, but significant contribution must
exist from immediate triggering by supernova explosions and the
radiation and stellar winds from massive stars
\citep[e.g.,][]{Peretto2012}. The filaments are expected to undergo
gravitational fragmentation that will lead to star formation
\citep[e.g.,][]{Inutsuka1997, Myers2009, Heitsch2011}. 

Many recent studies have addressed the question of the typical shape
of the filament cross sections. The observed column density profiles
are fitted with the so-called Plummer function, partly because of the
theoretical predictions associated with this functional form. In
particular, an infinite isothermal cylinder in hydrostatic equilibrium
is expected to follow this profile \citep{Stodolkiewicz1963,
Ostriker1964} while, for example, magnetic support, deviations from
the isothermal conditions, and initiated collapse would all cause
characteristic effects on the parameters of the profile function
\citep[see][]{Arzoumanian2011}. If the parameters of the profile
function can be determined with high accuracy, this will provide
important information on the formation and the current state of the
filaments. 

The central question is, can we accurately measure the structure and
mass of the filaments. Our knowledge is mostly based on molecular
lines and dust emission. The kinematic information provided by line
observations is crucial but the use of molecules for a structural
analysis is complicated by the abundance and excitation variations
that are difficult to measure and to model. The situation is
exacerbated in the dense environments where many molecules are
severely depleted as they begin to freeze onto dust grains.

Dust emission is considered a reliable probe of dense clouds,
especially at sub-millimetre wavelengths where the dependence on the
dust temperature is reduced. However, the line-of-sight temperature
variations can still play a significant role and, in the case of dense
cores, can render the column density estimates uncertain by a factor
of several \citep{Malinen2011}. Furthermore, dust is expected to
undergo evolution in dense and cold environments. This may include 
accretion of ice mantles, agglomeration of dust particles (leading to
the disappearance of small grains), and temperature-dependent changes
in the dust optical properties \citep[e.g.,][]{Ossenkopf1994,
Ormel2011, Meny2007, Kohler2011}. The analysis of sub-millimetre
observations is particularly affected by changes in the dust spectral
index $\beta$ and by the possible increase of the dust opacity
$\kappa$. The value of $\kappa$ may vary by a factor of several
between diffuse and dense regions. This is important for the study of
cloud structure, because the derived column densities are directly
proportional to $1/\kappa$. These effects could be expected to be
less severe in filaments than in the cloud cores because of the lower
density of the filaments and the consequently longer time scale of
grain growth. However, significant changes in dust sub-millimetre
opacity have been seen also towards filaments. For example, a factor
of $\sim$3.4 increase was reported for a filament in Taurus with
$A_{V}<10^{\rm m}$ \citep{Stepnik2003}. In addition to the uncertainty
of $\kappa$, a wrong assumption of $\beta$ will bias the mass
estimates through its effect on the colour temperature estimates
\citep{Malinen2011, JuvelaYsard2012, YsardJuvela2012}.

In this paper we investigate the possible differences between the real
density structure of the filaments and the one deduced from typical
sub-millimetre observations of dust emission. The cloud models are
based on magnetohydrodynamic simulations with distinct filamentary
structures. The emphasis is not on the nature of the filaments
themselves but on the uncertainties of deriving their properties from
observations. We carry out radiative transfer modelling to produce
synthetic surface brightness maps and estimate the profile functions
of selected filaments. The results are compared to the real column
density structure read directly from the model.

The structure of the paper is the following. In
Sect.~\ref{sect:synthetic} we present the cloud models and the
calculations leading to synthetic surface brightness maps. In
Sect.~\ref{sect:methods} we describe the selected filaments and the
procedures used to fit them with Plummer profiles. The main results
are shown in Sect.~\ref{sect:results} where we list the results for
the filament profile fits, both using the true column density and the
column density estimated from the synthetic observations. We discuss
the results in Sect.~\ref{sect:discussion} before presenting our final
conclusions in Sect.~\ref{sect:conclusions}.

\begin{figure*}
\centering
\includegraphics[width=16.0cm]{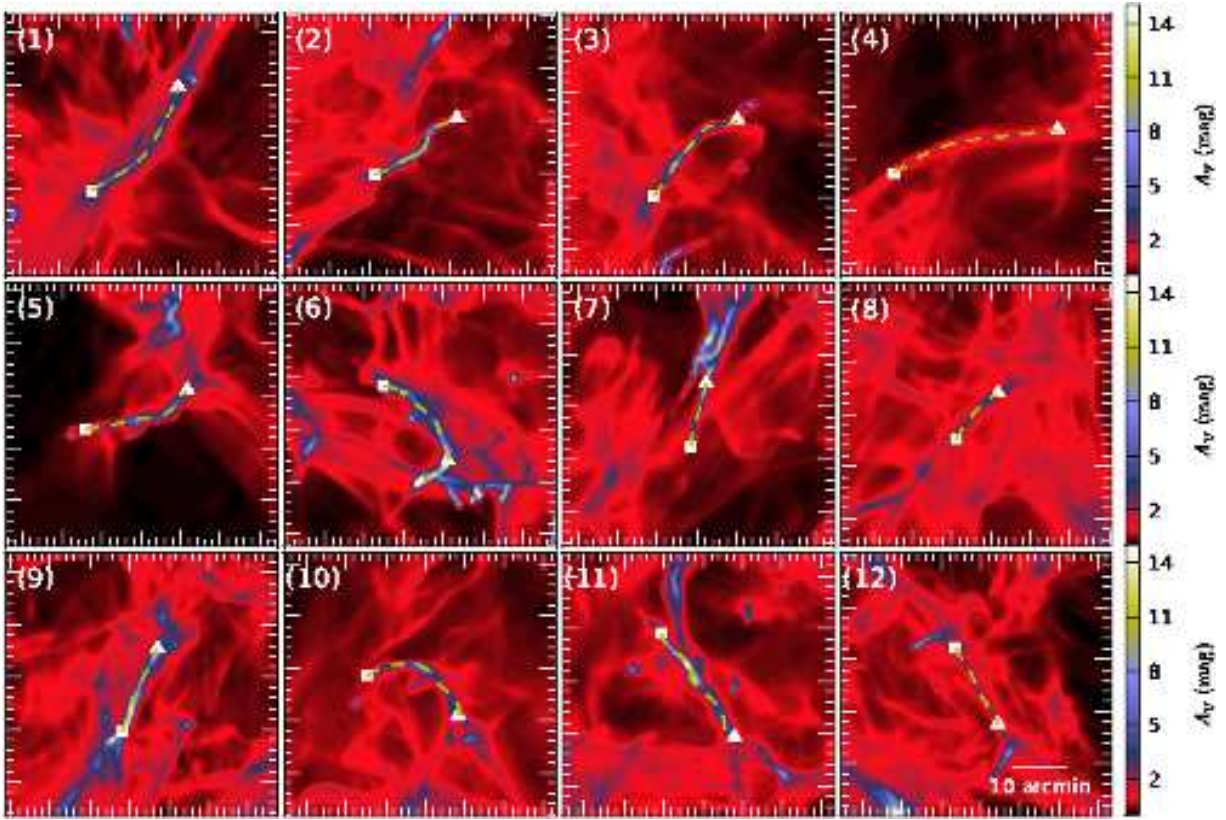}
\caption{
$A_{V}$ maps of the selected twelve filaments in model A. The
dashed lines indicate the filament segments included in the analysis.
The $A_{V}$ values are obtained directly from the model cloud
where the 1000$\times$1000 pixel image has been convolved with a
beam with FWHM equal to three pixels.
}
\label{fig:Av-maps}%
\end{figure*}

\begin{table}
\caption{Main parameters of the model clouds.}
\begin{tabular}{llll}
\hline \hline
Model  &   $<A_{V}>$   &  $<n($H$_2)>$  & Dust model  \\
       &   (mag)       &  (cm$^{-3}$)    &             \\
\hline       
A      &    1.0        &  111.0  &  standard MWD$^1$ \\
B      &    2.0        &  222.0  &  standard MWD  \\
C      &    4.0        &  450.0  &  standard MWD + dust with higher \\
       &               &         &  sub-mm opacity \\
\hline
\end{tabular}
$^1$\,Milky Way dust with $R_{V}=3.1$.
\end{table}

\section{Simulated clouds and observations} \label{sect:synthetic}

The modelling is based on a magnetohydrodynamic (MHD) simulation
carried out on a regular grid of 1000$^3$ cells. The rms sonic Mach
number is $\sim$9 and the rms Alfv\'enic Mach number $\sim$2.8. The
data are the same that were used in \citet{Malinen2011} and we refer
to that paper for details. 
The MHD calculations assume an isothermal equation of state. The
simulation was run first without gravity and the stirring continued
for several shock crossing times to develop turbulence well separated
from the initial conditions. The simulation was continued with gravity
switched on and the data used here corresponds to the situation during
early collapse \citep[Model I in][]{Malinen2011}.
We use the unigrid run~\citep{PadoanNordlund2011} rather than the
Adaptive Mesh Refinement (AMR) MHD data because we need high spatial
resolution not only for the cores but also for the lower density
filaments.

The cloud size was scaled so that the mean visual extinction $A_{V}$
through the cloud is 1.0$^{\rm m}$ or 2$^{\rm m}$. In the following,
these are called models A and B. Assuming a linear cloud size of
6\,pc, these correspond to mean hydrogen densities of 111\,cm$^{-3}$
and 222\,cm$^{-3}$, respectively. 

For the radiative transfer calculations we built an adaptive grid from
the $1000^3$ full resolution model. The adaptive grid is constructed
by combining neighbouring cells that are optically thin, and have
similar densities. The maximum allowed $V$ band optical thickness of a
cell is $\tau_{\mathrm{max}}=0.04$ in model A.
The maximum density contrast was set to
$\rho_{\mathrm{max}}/\rho_{\mathrm{min}}=30$. The resulting adaptive
model has approximately $160\times 10^6$ cells and
consists of
839 grids. 
The same grid was used for model B. The
gridding ensures that also moderately dense region, the filaments
in particular, are sampled at the best resolution. 

We use the radiative transfer program described in
\citet{Lunttila2012} \citep[see also][]{Juvela2003, Juvela2005}. The
cloud is illuminated externally by the isotropic interstellar
radiation field \citep{Mathis1983}. The dust is assumed to have the
properties of the standard Milky Way diffuse interstellar dust 
($R_{V}$=3.1) as described in \citet{Draine2003}. Because of the low
dust temperatures and the modest optical thickness of the model
clouds, the absorption of the thermal dust emission does not
contribute significantly to the dust heating. Therefore the dust
temperature was calculated in a single step without iteration.

In the third model C the average visual extinction is set to
$A_{V}\sim 4^{\rm m}$. Assuming a linear cloud size of 6\,pc, the mean
density is $n({\rm H})$=450\,cm$^{-3}$. In model C we use two dust
components. The second one corresponds to a modification where the
dust sub-millimetre opacity is increased. This is implemented by
multiplying the original opacities with a factor $(\lambda/20\mu{\rm
m})^{0.5}$ at wavelengths longer than 20\,$\mu$m. The resulting dust
opacity spectral index is thus $\beta\sim 1.5$ in the sub-millimetre
and, as a result, the sub-millimetre opacity $\kappa_{\nu}$ is higher.
For example at 250\,$\mu$m the $\kappa_{\nu}$ has increased by a factor of
3.5. The relative abundance of the two dust components is varied
according to the local density so that the abundances are equal at a
density of $n({\rm H})\sim 5\times 10^4$\,cm$^{-3}$ \citep[see Fig.5
in][]{Malinen2011}. At low densities the relative abundance of the
modified dust component goes to zero. By the time the density reaches
$10^6$\,cm$^{-3}$, practically all the dust is in the modified form.
The calculations for model C were carried out at a resolution of
500$^{3}$ cells where the grid was obtained by direct downsampling of
the original 1000$^{3}$ data.

The radiative transfer calculations \citep{Juvela2003, Lunttila2012}
solve the 3D temperature structure of the model clouds and result in
synthetic surface brightness maps of 1000$\times$1000 (models A and
B) or 500$\times$500 pixels (model C). Surface brightness maps
were calculated at 160, 250, 350, and 500\,$\mu$m that correspond to
the wavelengths of \emph{Herschel} observations. The
full-resolution images were convolved with Gaussian beams with FWHM
corresponding to 3, 6, or 12 pixels (in model C the beam widths
in pixels are half of these values). Before the convolution we
add to the maps white noise that corresponds to 3.7, 1.20, 0.85,
0.35\,MJy/sr per original \emph{Herschel} beam size of 12, 18, 25, and
37$\arcsec$, respectively. The values were obtained using the
\emph{Herschel} Interactive Processing Environment
HIPE\footnote{http://herschel.esac.esa.int/HIPE\_download.shtml} and
they correspond to the instrumental noise levels of typical \emph{Herschel}
mapping observations \citep[as in][]{Malinen2011}.
Most actual \emph{Herschel} observations are confusion limited, particularly
at wavelengths $\lambda \ge 250\,\mu$m. In our simulations the confusion
results only from structures inside the modelled volume. In more
realistic simulations, especially for low surface brightness values,
one might also have to consider the confusion by extragalactic point
sources. 
The final common resolution of all the
maps (i.e., the size of the beam used in the convolution) is
40$\arcsec$. 
When the data are convolved with a beam equal to three pixels,
the pixel size is thus 13.3$\arcsec$ and the map size is 3.7 degrees.
With the assumed cloud size of 6\,pc, the FWHM beam sizes of 3,
6, and 12 pixels translate to the cloud distances of 93, 186, and 371\,pc
(in the following these are called the near, medium, and far
distances). For the near distance, the filaments are clearly
larger than the beam. For the medium distance, the filaments are
still slightly larger than the beam, but in the case of the far
distance, the observed filament width is dominated by the convolution.

\begin{figure}
\centering
\includegraphics[width=7.0cm]{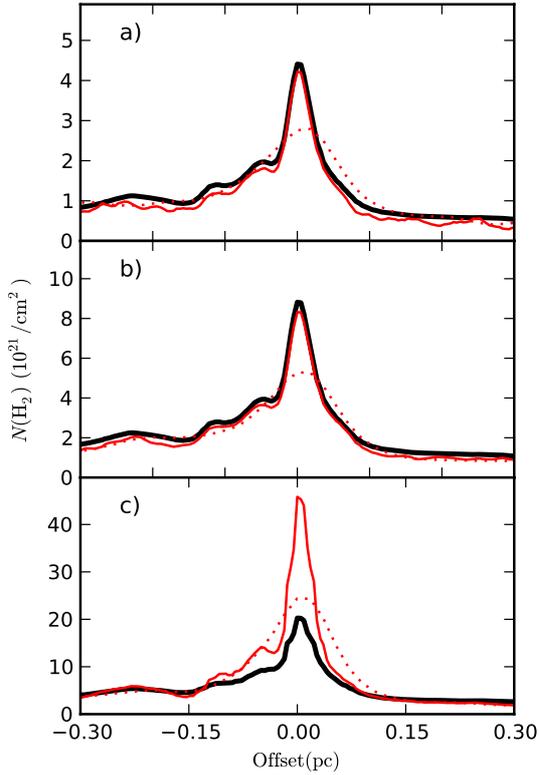}
\caption{
The profile of the filament number 1 as seen in the models A, B,
and C (frames $a$, $b$, and $c$, respectively). The solid thick line
shows the actual column density profile in the model cloud. The other
curves show the profiles derived from the synthetic observations when
the cloud is at a near distance of 93\,pc (thin solid line)
or 371\,pc (dotted line).
}
\label{fig:profile_examples}%
\end{figure}

\section{Methods of analysis}  \label{sect:methods}

The synthetic surface brightness observations were converted to maps
of column density. The dust colour temperature was calculated pixel by
pixel by fitting the observed 160, 250, 350, and 500\,$\mu$m
intensities with a modified black body curve. The data points were
weighted according to the noise levels of the simulation. In the fits
the opacity spectral index was fixed to a constant value of
$\beta=1.8$ or $\beta=2.0$. These values differ from the actual value
in the dust model (Milky Way dust with $R_{V}=3.1$) where the
spectral index between the wavelengths of 160\,$\mu$m and 500\,$\mu$m
is 2.09.

With the dust colour temperature known, the column density is
calculated as 
\begin{equation}
   N^{\rm Obs} = \frac{ S_{\nu} } {B_{\nu}(T_{\rm C})  \kappa_{\nu}},
\end{equation}
where $S_{\nu}$ is the intensity of the fitted modified black body
curve, $B_{\nu}$ is the Planck function, and $\kappa_{\nu}$ the dust
opacity. The $\kappa_{\nu}$ value was taken directly from the dust
model. Therefore, the derived column density $N^{\rm Obs}$ differs
from the true column density $N^{\rm True}$ only because of noise and
because of systematic errors in the colour temperature $T_{\rm C}$,
the latter being caused by line-of-sight temperature variations.

We selected four filaments from each of the three column density
images calculated towards three orthogonal directions (see
Fig.~\ref{fig:Av-maps}). The filament segments were selected by eye
and the path of each filament was set to follow the peak of the
filament as seen in the $N^{\rm True}$ maps obtained from the cloud
model.

The procedure used in analysing the filaments is basically the same as
in \citet{PaperIII}. The column density profiles perpendicular to the
filament path were extracted at 20$\arcsec$ intervals. The individual
profiles were aligned so that their peaks match before calculating the
median profile of the whole filament. Thus the location of the
filament was allowed to differ from the originally selected path. A
fixed range of [$-$0.3\,pc, $+$0.3\,pc] around the peak of the column
density profile was included in the subsequent analysis. The filament FWHM was
measured by fitting the profile with a Gaussian function plus a
constant background. The filament FWHM values were determined for each cross
section separately and for the final combined median profile. The
median profile of each filament was also fitted with a model
consisting of a linear background plus a Plummer-like column density
profile
\begin{equation}
N(r)  = A_{p} \frac{\rho_{\rm c}R_{\rm flat}}{[1+(r/R_{\rm
flat})^2]^{(p-1)/2}}
\end{equation}
\citep[see][]{Arzoumanian2011, Nutter2008, PaperIII}.
The factor $A_{p}$ is calculated using the formula
$A_{p} = \int_{-\infty}^{\infty}
[  (1+u^2)^{p/2}  \, \, {\rm cos} \, i]^{-1} \, du$,
where we assume an inclination angle of $i=0$ \citep{Arzoumanian2011}.
In addition to the two parameters describing the background, the free parameters of
the Plummer function are the central density of the filament
$\rho_{\rm c}$, the size of the inner flat part $R_{\rm flat}$, and
the power-law exponent $p$.
The finite resolution of the observations was taken into account by
fitting the observed profile not directly with the theoretical Plummer
profile but to a Plummer profile convolved with the beam used in
the simulated observations.
The convolution weights take into account the fact that the actual
convolution takes place over a 2D map. This allows us to
derive deconvolved parameters for even unresolved filaments.

In addition to the fitted parameters, we also look at some derived
quantities. In particular, the mass per unit length of filament,
$M_{\rm line}$, can be calculated by integrating over the fitted
Plummer profile.

\section{Results} \label{sect:results}  

The twelve filaments were analysed using the true column density
$N^{\rm True}$ and the column densities $N^{\rm Obs}$ estimated from
synthetic observations. The analysis was repeated with an actual
resolution equal to 3, 6, and 12 pixels of the original
1000$\times$1000 pixel maps, these corresponding to the near,
medium, and far cloud distances. Because of the lower resolution of
the model C, these correspond there to convolution with a beam
with FWHM equal to 1.5, 3, and 6 pixels, respectively.

Figure~\ref{fig:profile_examples} shows examples of the column density
profiles, comparing the true column density profiles to those estimated
from the synthetic observations. These already indicate that in models
A and B the column densities are underestimated by $\sim$10\% at the centre of the filament,
because the estimated colour temperature $T_{\rm C}$ overestimates
the mass-averaged dust temperature. The fact that the analysis is
carried out with a slightly too low value of the spectral index,
$\beta=2.0$, instead of 2.09, contributes in the same direction but
is a smaller factor. In model C, the sub-millimetre dust
opacity was increased by a factor of a few in the densest parts of the
cloud. This is directly reflected in the column density estimates that
are more than twice the correct value at the centre of the filaments.
The figure also shows that, apart from the filament itself, the
profiles are affected by other emission components, i.e., by background
confusion.

\begin{figure*}
\centering
\includegraphics[width=16.6cm]{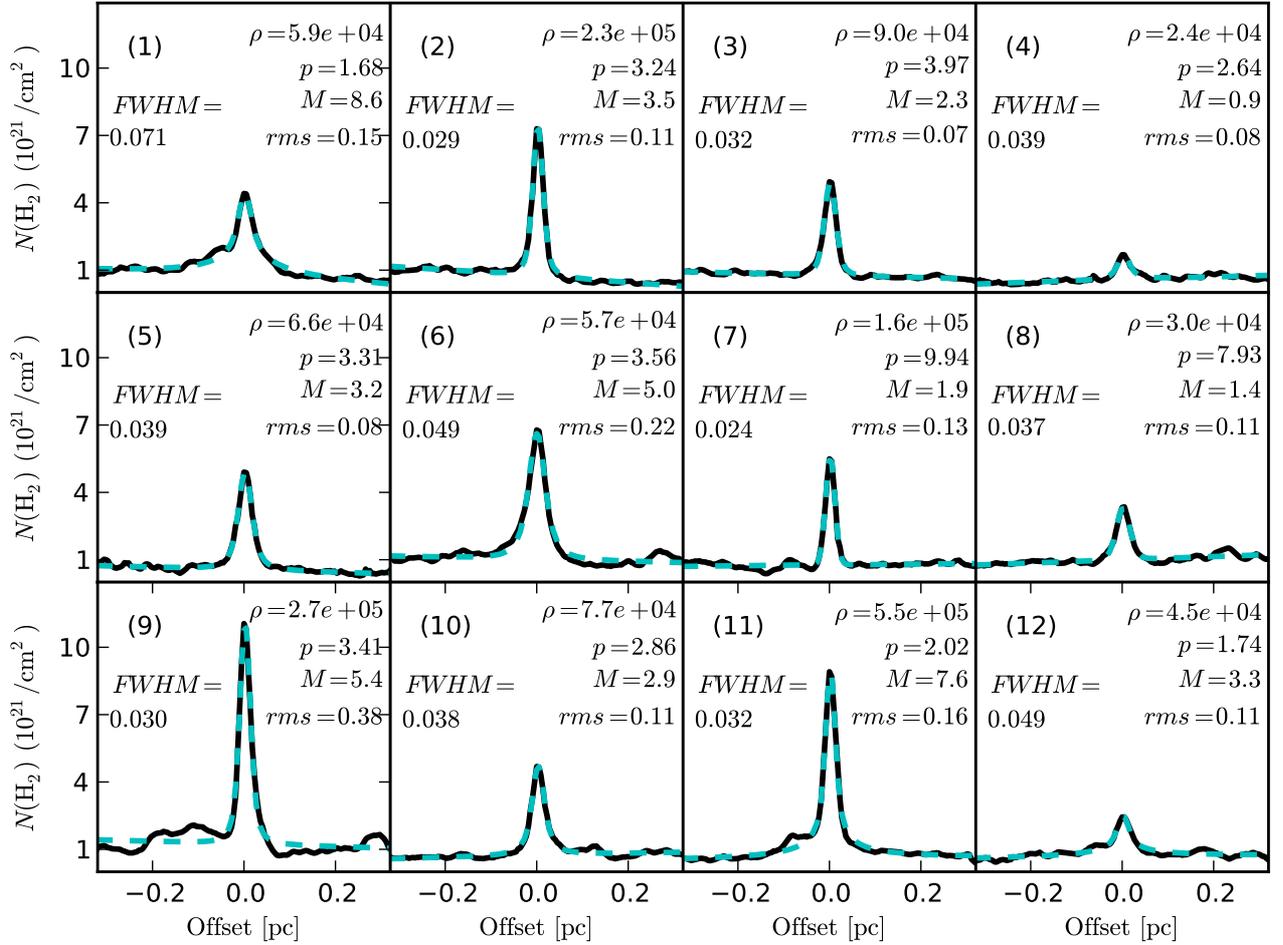}
\caption{
Fits of the Plummer function to profiles extracted from the
observed column density maps ($\beta$=2.0) of Model A cloud at the
near distance. The solid black curves show the estimated median
profiles of the filaments and the dashed cyan lines the fitted
functions. The parameters $\rho_{\rm c}$ (central number density of
H$_2$ in units of cm$^{-3}$), $p$, and $M_{\rm line}$ (in units of
$M_{\sun}$/pc) are shown in the frames as well as the rms value of the
fit (in units of $10^{21}$\,cm$^{-2}$) and the FWHM values of the
filaments obtained from fits with Gaussian functions.
}
\label{fig:A_fwhm3}%
\end{figure*}

Each filament was fitted with a Plummer function.
Fig.~\ref{fig:A_fwhm3} shows the results for model A in the case
where the observed column densities were estimated with $\beta=2.0$
and the cloud distance was set to $d=93$\,pc, which corresponds to
convolution with a beam with the FWHM equal to three pixels. The
figure demonstrates that the profiles can be well fitted using the
Plummer function.
In Fig.~\ref{fig:A_fwhm3} the parameter values show a large scatter.
\citet{Malinen2012} noted that the fitting of Plummer profiles is
sometimes sensitive to the initial parameter values of the least
squares minimisation. Visual inspection shows that all the fits in
Fig.~\ref{fig:A_fwhm3} are good. However, in addition to intrinsic
variations between the filaments, some of the scatter is caused by the
instrumental noise and especially by the confusion caused by other
emission along the line-of-sight.

\begin{figure*}
\centering
\includegraphics[width=13cm]{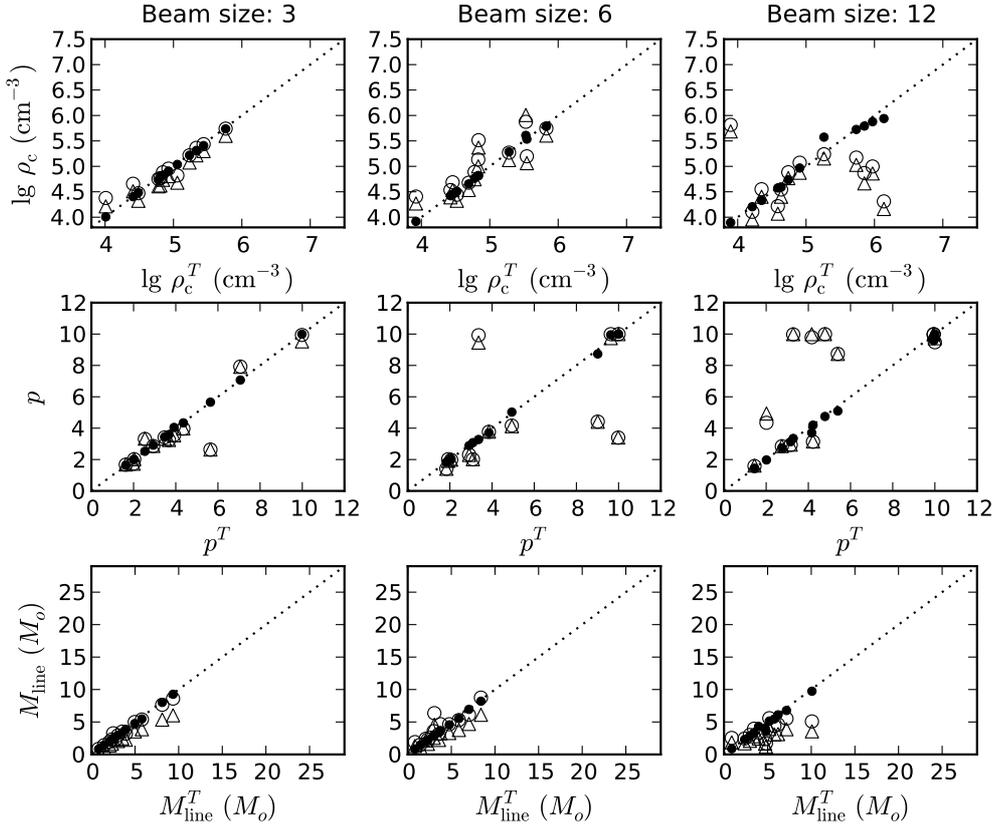}
\caption{
Comparison of the Plummer parameters derived from true column density
and the values estimated from the synthetic maps of model A. The rows
correspond to parameters $\rho_{\rm c}$, $p$, and $M_{\rm line}$. 
The columns correspond to beam sizes equal to 3, 6, and 12 pixels,
i.e., to the near, medium and far cloud distances. On the $x$-axis
are always the parameter values from $N^{\rm True}$ and on the
$y$-axis are the values from $N^{\rm Obs}$ derived with $\beta=2.0$
and without observational noise (solid circles), with $\beta=2.0$ and
noise (open circles), and with $\beta=1.8$ and noise (triangles).
}
\label{fig:compare_A}%
\end{figure*}

Figure~\ref{fig:compare_A} shows a comparison of the estimated values
of $\rho_{\rm c}$, $p$, and $M_{\rm line}$ for all three beam
sizes (3, 6, and 12 pixels) for model A. On the $x$ axis are
the parameters obtained using the true column densities, on the $y$
axis are the values derived from the analysis of synthetic
observations with and without observational noise and with $\beta$
equal to 1.8 or 2.0. The effect of the $\Delta \beta=0.2$ is
relatively small and is seen mainly as a systematic effect in the
density $\rho_{\rm c}$ and mass $M_{\rm Line}$ values where it amounts
to an underestimation of 20--30\%. Without noise the parameter values
obtained using observed and true column density are similar to within
a few percent.  
The effect of noise is often small but can also lead to tens of
percent difference in the parameter values. The loss of resolution
(i.e., convolution with a larger beam) has a clear impact when noise
is included and leads to larger scatter especially in the estimates of
the central density, which are underestimated in most cases, even up
to $\sim$2 orders of magnitude. In the fitting routine, the maximum
allowed value for the parameter $p$ is 10. The largest bias in $p$
values occurs when the value derived from either the true or observed
column density (but not both) is $\sim$10.  
Compared to the other parameters (especially the central density
$\rho_{\rm c}$), the mass $M_{\rm Line}$ is less affected by the noise
and the resolution.

\begin{figure*}
\centering
\includegraphics[width=13cm]{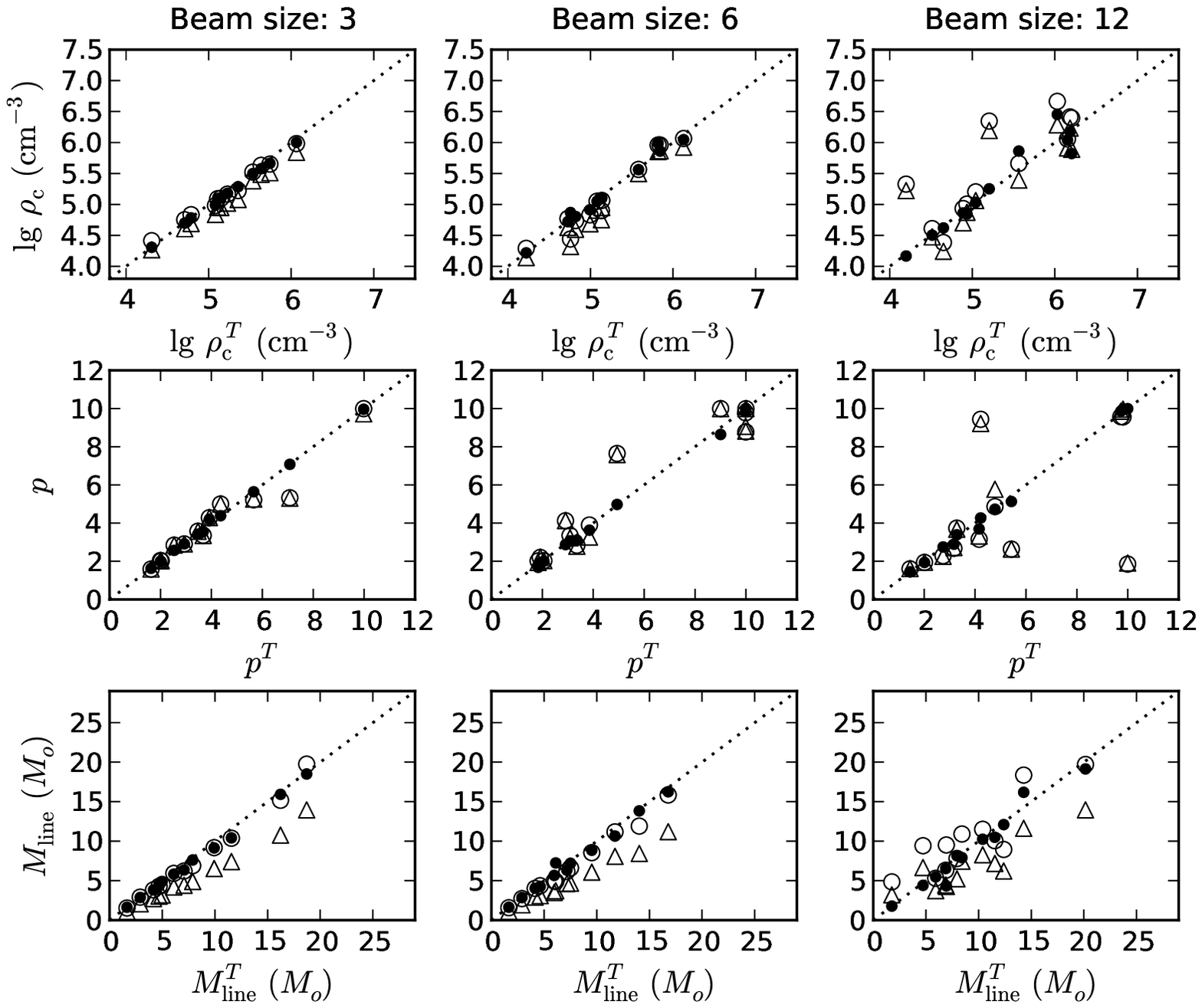}
\caption{
As Fig.~\ref{fig:compare_A} but for the model B that has twice as high
column density.
}
\label{fig:compare_B}%
\end{figure*}

The corresponding results for the higher column density model B are
presented in Fig.~\ref{fig:compare_B}.  Also here the parameter values
obtained using observed and true column density without 
instrumental noise are in most cases similar to within a few percent.

The scatter of the central density $\rho_{\rm c}$ values (from $N^{\rm
Obs}$ including noise vs. those from $N^{\rm True}$) appears to be
lower compared to model A. 
Using the observations with beam sizes equal to 3 and 6 pixels
(corresponding to the near and medium cloud distances) the density and
mass are mostly either correct or somewhat underestimated. However,
when low resolution with a beam size of 12 pixels is used, the
scatter increases and the parameters can be also overestimated. 
For the $M_{\rm Line}$ values the error is up to a factor of $\sim$2. 
One should note that this is the
error between the masses derived from either the true or the estimated
column density. We will return later to the comparison with the actual
mass of the filaments that can be obtained directly from the model cloud.

\begin{figure*}
\centering
\includegraphics[width=16cm]{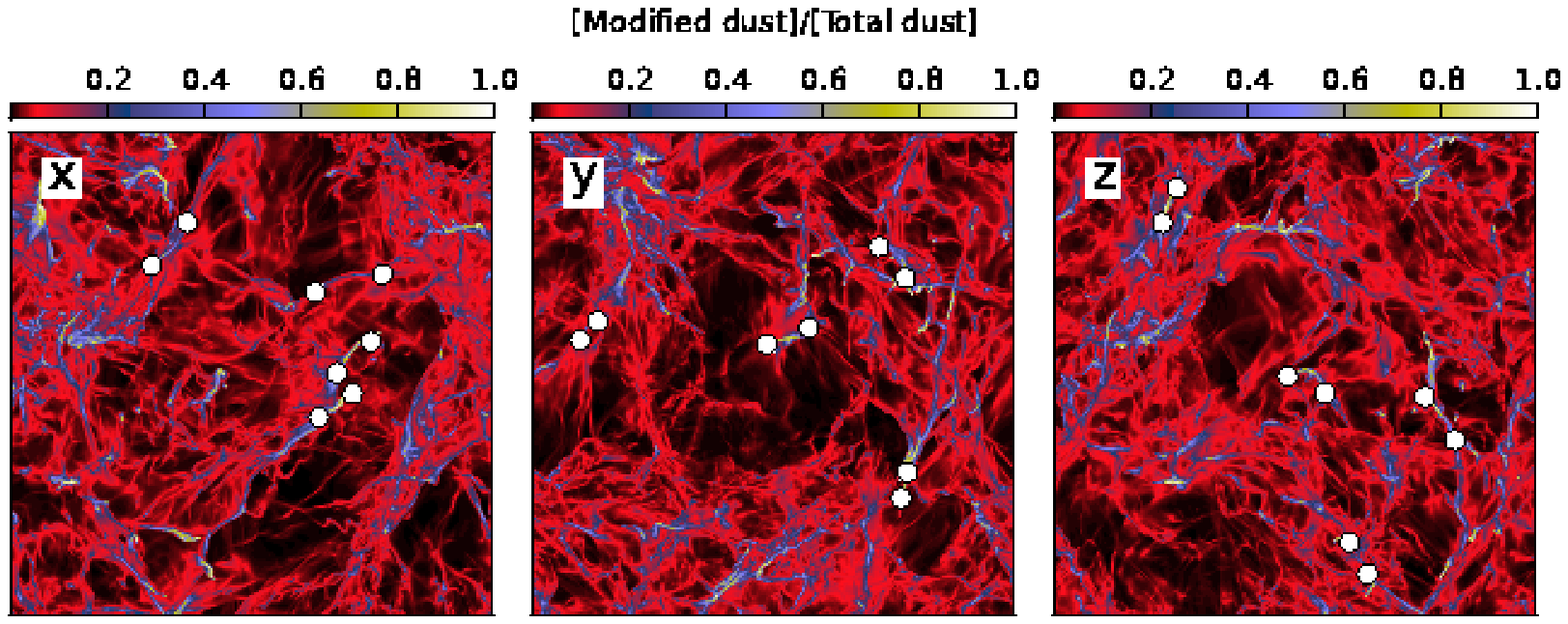}
\caption{
The fractional abundance of modified dust in the model C. The three
frames correspond to three orthogonal directions of observations. The
white circles indicate the ends of the filament segments included in
the analysis. 
}
\label{fig:abu_C}%
\end{figure*}

In model C, the abundance of modified dust, i.e., the component with
higher sub-millimetre opacity, increases with density.
Figure~\ref{fig:abu_C} shows the relative abundance of the modified
dust as projected values towards three orthogonal directions. The locations
of the selected filaments are marked in the figure. The figure
shows that while modified dust is the dominant component towards the
centre of many filaments, its abundance remains below 20\% in most of
the area included in the Plummer fits, i.e., at distances of a few
arcmin from the filament centre. Thus, the spatial variation of dust
properties should be able to bias both the $N_{\rm Obs}$ values and
the observed shape of the profiles and thus affect the parameters of
the Plummer fits.

\begin{figure}
\centering
\includegraphics[width=8.3cm]{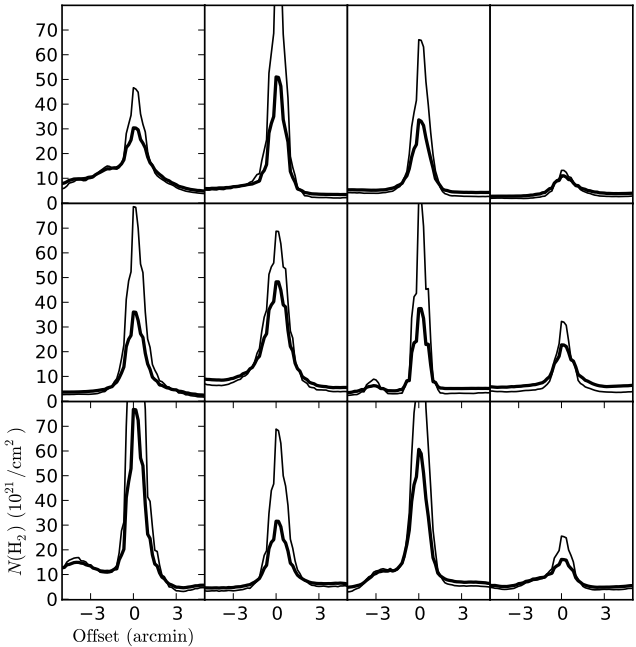}
\caption{
Profiles of the filaments in model C (at the near cloud
distance). The thick lines correspond to the true column density
$N^{\rm True}$ of the model cloud multiplied by a factor of 1.5. The
thin lines denote the column density derived from synthetic
observations. 
}
\label{fig:model_C_profiles}%
\end{figure}

Figure~\ref{fig:model_C_profiles} presents some examples of the column
density profiles of the model C filaments, comparing $N^{\rm Obs}$
and $N^{\rm True}$. Because the column densities are estimated
assuming a constant dust opacity, $\kappa_{\nu}$, the column densities
are overestimated whenever the dust opacity has been increased. In the
figure the $N^{\rm True}$ values have been scaled by a factor of 1.5.
This provides rough agreement with $N^{\rm Obs}$ in the outer parts of
the filaments, while at the centre the synthetic observations lead to
still higher column densities by a factor of $\sim$2. For example, the
effect on the observed filament FWHM depends on whether the
emission at the half value point is produced mainly by the normal dust
or the modified dust. In the latter case, the observed filament
FWHM would not be affected even though the column density can be
strongly overestimated.

\begin{figure*}
\centering
\includegraphics[width=14cm]{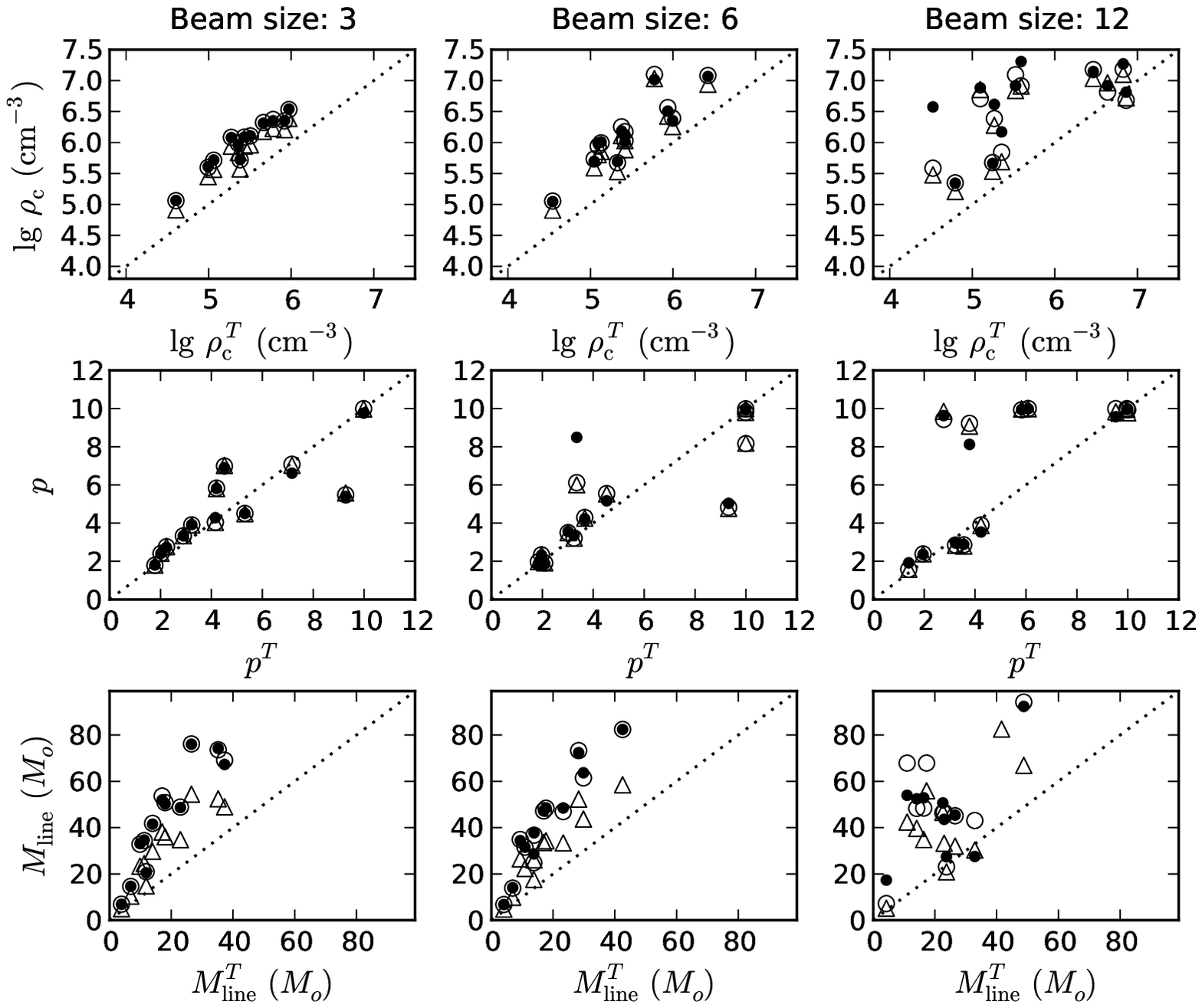}
\caption{
As Fig.~\ref{fig:compare_A} but for the model C containing two dust
components with a spatially varying abundance ratio. The beam
size (3, 6, or 12 pixels corresponding to the near, medium, or far
cloud distance, respectively) is indicated on top of each column of
frames.
}
\label{fig:compare_C}
\end{figure*}

The values of the Plummer parameters obtained with model C are shown
in Fig.~\ref{fig:compare_C}. The optical depth of the cloud is about
twice as high as in the case of model B. Therefore, in addition to
the spatial variation of dust properties, also the radiative transfer
effects should have a stronger effect on the results. The observations
are seen to overestimate the filament mass up to a factor of several.
Thus, the increase of the average $\kappa_{\nu}$ dominates over the
opposite effect resulting from the dispersion of dust temperatures
along the line of sight. The volume density estimates can be
$\sim$0.5--1 orders of magnitude too high and are more biased when the
resolution of the observations is low. The scatter is larger than in
the previous models, especially when the data are convolved with the
largest 12 pixel beam. 
The smaller spectral index $\beta$=1.8 still leads to smaller $M_{\rm
Line}$ values but the scatter of the values suggests that the fits are
less well constrained. The $p$ values can be divided to two groups:
those close to $p \sim$10 and those with values $p \sim$2--4. At low
$p$ the values obtained from synthetic observations do not show any
systematic bias. Contrary to models A and B, in model C there
are also cases without instrumental noise where the $p$ value
derived from either the observed column density or the true column
density (but not both) is close to 10.

\begin{figure}
\centering
\includegraphics[width=7.9cm]{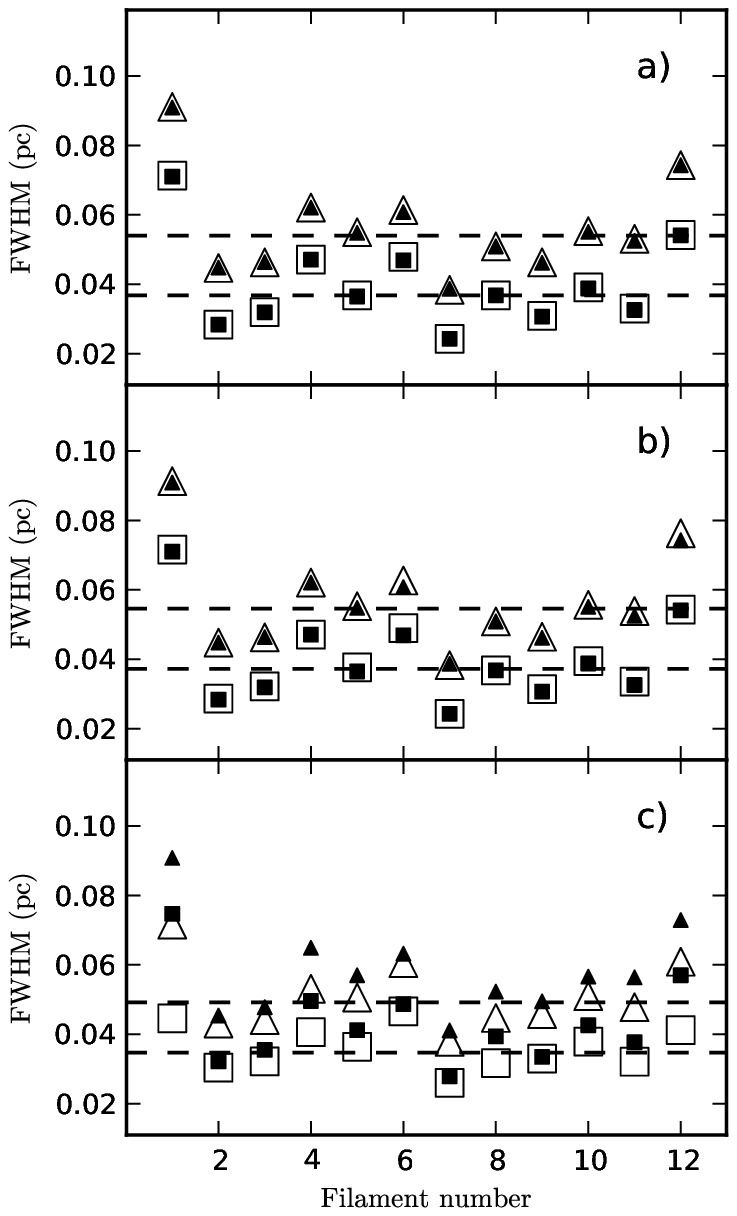}
\caption{
The FWHM values of the twelve filaments. The frames $a$, $b$, and $c$
correspond to the models A, B, and C, respectively. The squares
and the triangles correspond to the near cloud distance of
93\,pc and the medium cloud distance of 186\,pc, respectively.
The solid symbols are values derived from $N^{\rm True}$ and the open
symbols values from $N^{\rm Obs}$. The dashed lines indicate the
median values for the FWHM values derived from $N^{\rm Obs}$. Upper
line corresponds to the distance of 186\,pc and lower line to the
distance of 93\,pc.
}
\label{fig:fwhm}
\end{figure}

The FWHM values of the filaments were calculated with Gaussian fits
and Fig.~\ref{fig:fwhm} compares the values in the three models. The
values derived from $N^{\rm True}$ should be identical in all the
three models. However, because of the coarser initial resolution of
model C, the values of that model differ slightly. In models A and
B the values obtained from synthetic observations are always within
a couple of per cent of the $N^{\rm True}$ results. In model B the
filament optical depths are above $A_{V}=10^{\rm m}$ and one could
expect the central column density to be underestimated relative to the
lower density outer regions. However, the effect on the beam
convolution is still almost negligible. More dispersion is seen in
model C. Because the enhanced emissivity increases the signal at the
centre of the filaments, the estimated FWHM values of the
filaments would be expected to decrease. This is also the observed
trend, the analysis of $N^{\rm Obs}$ leading to $\sim 10-20$\% lower
values than $N^{\rm True}$. The effect is not quite constant, probably
because of the additional effect of instrumental noise and the
confusion by dense background structures. The largest differences are
$\sim$40\% (filaments number 1 and 12).

\begin{figure}
\centering
\includegraphics[width=6.5cm]{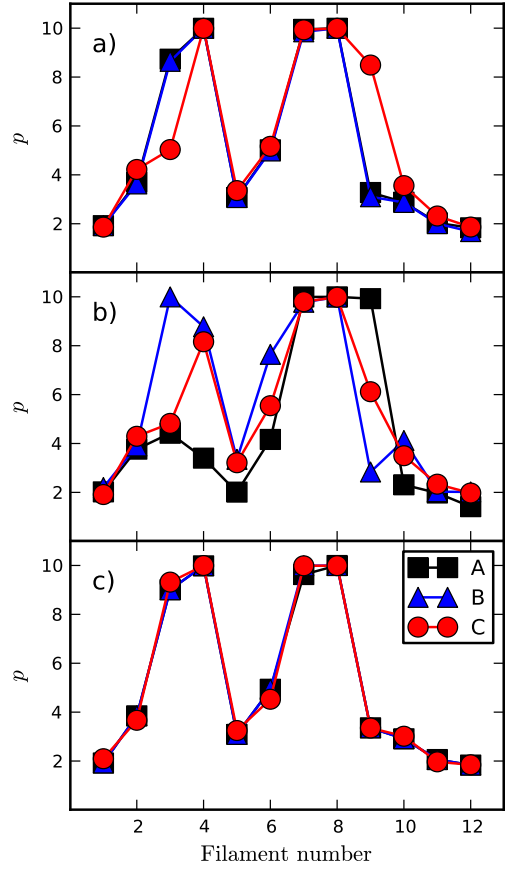}
\caption{
The $p$ values of the twelve filaments using medium cloud
distance of 186\,pc (beam size of 6 pixels) and $\beta=$2.0.
Squares indicate Model A, triangles Model B, and circles Model
C. $a$) Values are derived from the observed $N^{\rm Obs}$ without
noise. $b$) Values are derived from the observed $N^{\rm Obs}$ with
noise. $c$) Values are derived from the true $N^{\rm True}$.
}
\label{fig:ps}
\end{figure}

We compare the $p$ values of the twelve filaments in the three models
in Fig.~\ref{fig:ps}, using the medium cloud distance of 186\,pc
(beam size equal to 6 pixels) and $\beta=2.0$. When the
parameter values are derived from the true column density, $N^{\rm
True}$, (frame $c$), all the models give similar results regardless of
the value. In four out of twelve cases the $p$ value is above 8,
indicating that the parameters are not well constrained, i.e., the
minimum $\chi^2$ value is reached with parameter values that are far
from those obtained in the absence of noise. Most of the other twelve
filaments have $p$ values in the range 2--4. When the parameter values
are derived from the observed $N^{\rm Obs}$ without instrumental
noise (frame $a$), the results are mostly similar (to within $\Delta p
\sim$ 0.5) to the parameter values derived from the true $N^{\rm
True}$. Only in two cases the $p$ value in model C is biased by a
factor of $\sim$2, notably to opposite directions. However, when using
the observed $N^{\rm Obs}$ with noise (frame $b$), the situation gets
more complicated. Now only half of the filaments have similar $p$
values (to within $\Delta p \sim$ 0.5) as in the case with the true
$N^{\rm True}$. This applies to all three models. The bias can now
also be seen in all models, in particular in model A. These
results indicate that the effect of instrumental noise on the derived
$p$ parameter is more prominent than the radiative transfer effects.

\begin{figure}
\centering
\includegraphics[width=7.9cm]{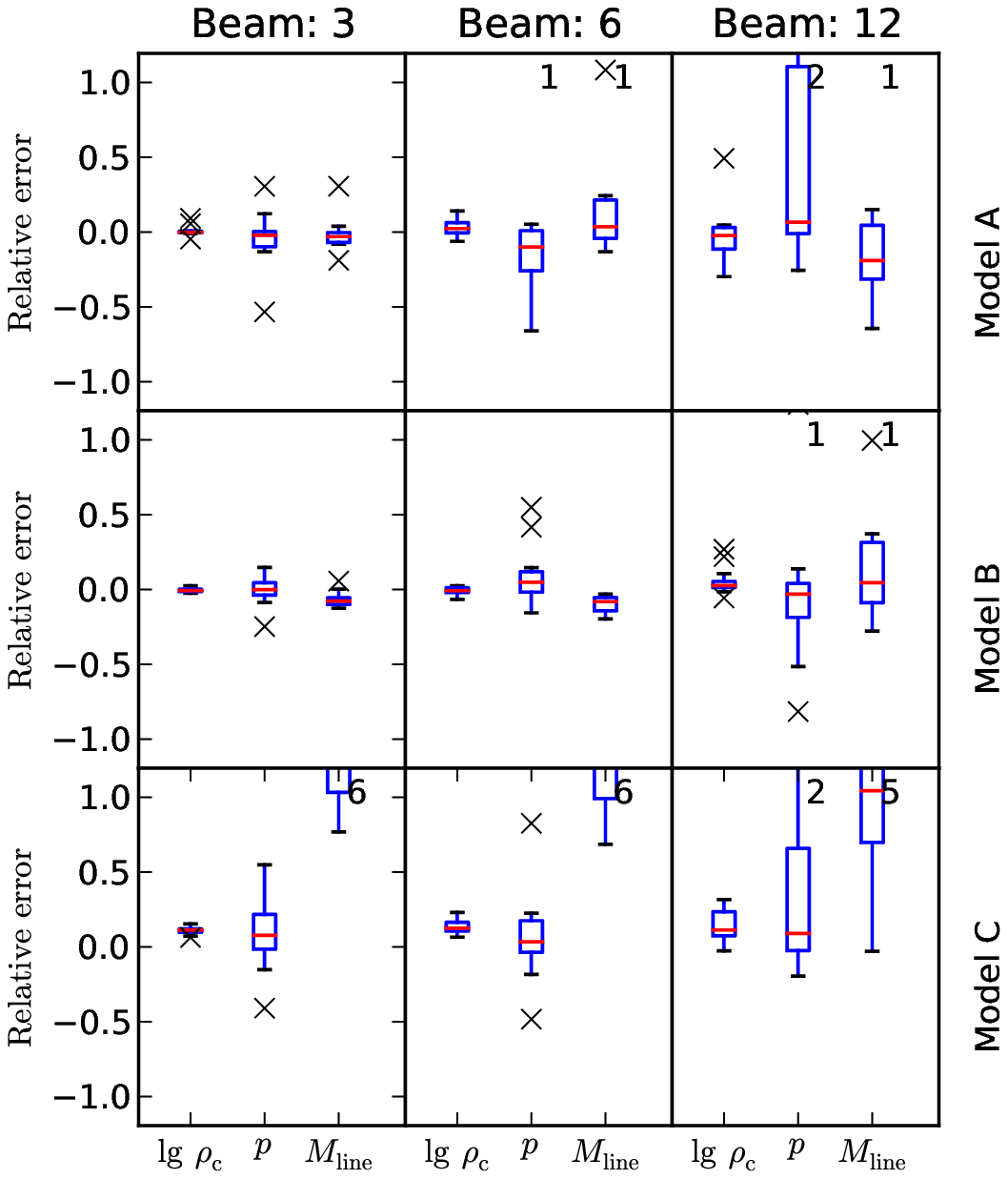}
\caption{
Boxplot for the relative errors of lg$\,\rho_{\rm c}$, $p$, and
$M_{\rm line}$ for the three cloud models and the three cloud
distances (beam sizes of 3, 6, and 12 pixels corresponding to the
near, medium, and far distances). The errors are calculated by 
comparing the values derived from $N^{\rm Obs}$ to those obtained from
$N^{\rm True}$ (see Figs.~\ref{fig:compare_A}, \ref{fig:compare_B},
and \ref{fig:compare_C}). For each parameter, the box extends between
the first and third quartile with a horizontal line indicating the
median value. The vertical lines (whiskers) extend from the box to the
last data point within a distance equal to 1.5 times the distance
between the third and the first quartile. Points outside this range
are drawn with crosses. The number of outliers falling outside the
plotted range is indicated with numbers near the upper and lower
borders.
}
\label{fig:boxplot}
\end{figure}

The results of Figs.~\ref{fig:compare_A}, ~\ref{fig:compare_B}, and 
~\ref{fig:compare_C}, are summarised in Fig.~\ref{fig:boxplot}. The
figure shows for the three model clouds and the three cloud distances
the distributions of the relative parameters errors using 
boxplots\footnote{See {\tt http://en.wikipedia.org/wiki/Box\_plot}}.
The errors are calculated as the difference between the values derived
from $N^{\rm Obs}$ (with $\beta$=2.0) and $N^{\rm True}$, normalised
with the latter parameter estimate. The figure includes the error
distributions for the logarithm of the central density $\rho_{\rm c}$,
parameter $p$, and the mass $M_{\rm line}$.

\section{Discussion} \label{sect:discussion}  

The estimated central density of the filament $\rho_{\rm c}$, the
power-law exponent $p$, and the mass per length unit $M_{\rm Line}$
were examined in Sect.~\ref{sect:results}. 
We have shown that the filaments of our MHD model can be well fitted
with the Plummer function. However, the fitting is sometimes sensitive
to the initial values of the fitting parameters (not all initial
values leading to a good fit) and the parameters extracted from
apparently good fits can have a large scatter, especially in the
case of filaments that are not well resolved. This indicates that the
parameters are sometimes sensitive to the amount of instrumental noise
added and thus very likely also to the confusion by background
structures. All the fits used in this paper look visually good, i.e.,
the least squares procedure has converged to an apparently valid
solution. However, different initial values might lead to slightly
different combinations of parameters values because, for example,
$R_{\rm flat}$ and $p$, are intrinsically correlated.
\citet{Malinen2012} also noted that small changes in the input data
can sometimes have a large effect on the estimated Plummer parameters.

In models A and B, when observational (i.e., instrumental)
noise was not included, the parameters derived from $N^{\rm True}$ and
$N^{\rm Obs}$ were similar to within a few percent. 
Furthermore, in the case of the near distance cloud (beam
size equal to 3 pixels) the results were not very sensitive to
the level of instrumental noise. When the resolution of the
observations was poor, in particular for the far cloud distance
of 371\,pc that corresponded to convolution with a beam with
FWHM equal to 12 pixels, noise did cause significant scatter. In the
central density $\rho_{\rm c}$ the errors could occasionally increase
up to two orders of magnitude. 
In this case the beam is larger than the width of the filament
and the observations give little constraints on the central profile.
Occasionally the result is a very narrow and high density peak that,
when convolved with the large beam, is still consistent with the
observations.

Compared to the model A, the model B was much less affected by the
noise because of its higher column density and thus higher
signal-to-noise ratios. In model C, with modified dust, central
density was overestimated by up to $\sim$0.5--1 orders of magnitude,
or even more with the lowest resolution. Similarly, the mass estimates
could be in error by a factor of several. The scatter was much larger
than in the models A and B, especially for the medium and far
cloud distances.  For the clouds at the far distance (lowest
resolution), the total width of the profile is dominated by the
convolution and the values of the derived Plummer parameters are
more uncertain. 
This is clearly demonstrated by Fig.~\ref{fig:boxplot}. The typical
errors increase by a factor of several between the near and far
distances. On the other hand, in the case of the model B the errors
are only half of those found in the model A. For model C the errors
are larger. The mass estimates are directly affected by the wrong
assumption of dust opacity but the other parameters do not show 
very significant bias.

The central question is, are the fitted parameter values, in particular
$p$, so reliable that theoretical conclusions can be made based on
them. In all our models the $p$ parameter values have a bipolar
behaviour. The lower values ($\sim$2--4) are mostly within a few
percent from those derived from the true column density. The largest
errors occur when the $p$ value derived from either the observed or
true column density is near the allowed maximum value of 10. In models
A and B this occurs only when instrumental noise is
included, but in model C also without noise. Noise and low
resolution clearly increase the probability of getting a high value
for $p$, which implies that the parameters are not well constrained.
\citet{Malinen2012} have shown that the confidence regions of the Plummer
parameters can be wide and the parameters (e.g., $p$ and $R_{\rm
flat}$) strongly correlated. We also compared the $p$ parameter values
derived from our three models for individual filaments
(Fig.~\ref{fig:ps}). The effect of noise appears to be much stronger
than the radiative transfer effects.

\begin{figure}
\centering
\includegraphics[width=6.5cm]{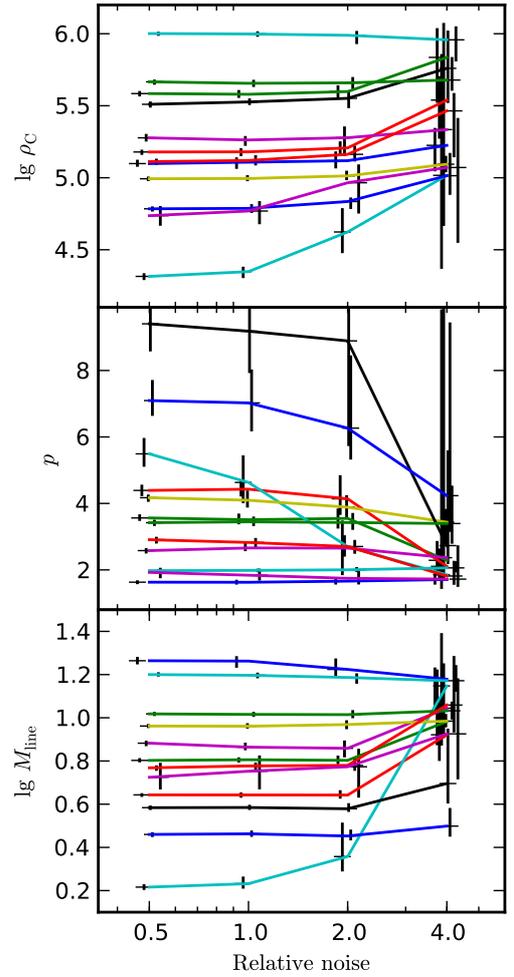}
\caption{
The dependence of the extracted Plummer parameters $\rho_{\rm c}$,
$p$, and $M_{\rm line}$ on the noise level in the case of model A and
the near cloud distance. The level of instrumental noise is either
0.5, 1.0, 2.0, or 4.0 times the default value (see
Sect.~\ref{sect:synthetic}). The plot shows for all the twelve
filaments the median values and, as vertical bars, the inter-quartile
intervals of the estimated parameters that are obtained from 50 noise
realisations of the surface brightness maps. The coloured lines
connect the median values. The vertical bars plotted for the
individual filaments are shifted slightly along the $x$ axis to avoid
overlap.
}
\label{fig:para_vs_noise}
\end{figure}

To separate the effects of instrumental noise from those of spatial
resolution we carried out tests where only the noise levels were
varied. Figure~\ref{fig:para_vs_noise} shows the results for model A
at the near distance. The noise is scaled between 0.5 and 4.0 times
the values used above. The figure shows the median parameter values
and their inter-quartile ranges for each filament and each noise
level, as determined from 50 noise realisations. With the default
noise, all the parameters shown are still well determined. When the
noise is increased, the scatter between noise realisations increases
rapidly. The noise also causes bias in the case of some filaments.
However, this varies from filament to filament and, for example, the
median value of $M_{\rm line}$ can either increase or decrease. The
figure also implies that as the cloud distance is increased in the
simulations, part of the increased scatter in the parameter values is
caused by noise. At the far distance, the filament appears four time
smaller than at the near distance and the total signal-to-noise
ratio (integrated over the source) is correspondingly smaller. At the
relative noise value of 4.0 the value of $p$ is well constrained
(within 50\%) for less than half of the selected filaments. In the
fits $p$ was always restricted to values below 10.0.

We have concentrated on the differences between the true and the
observed column densities and their effect on the estimated filament
parameters. However, we can note some additional factors affecting the
interpretation of such parameters. Like in our models, the
line-of-sight confusion can be a problem in real observations. For
example, some of the occasional large values of the parameter $p$ are
suspected to be caused by background structures. Those can be confused
with the filament itself or they may affect the fits by modifying the
background estimates (see Fig.~\ref{fig:A_fwhm3}).

\begin{figure}
\centering
\includegraphics[width=8cm]{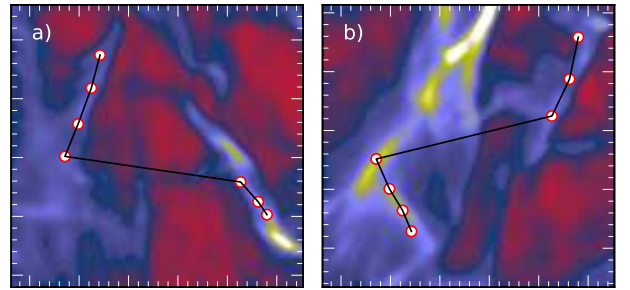}
\caption{
Filament 1 seen from two orthogonal viewing directions. The object was
identified from observations towards the $x$ direction. The figure
shows the object from the $y$ and $z$ directions (frames $a$ and $b$,
respectively). The circles denote volume density maxima for a few
lines of sight through the centre of the filament as seen towards the $x$
direction. The background image is the $A_{V}$ of the model cloud.
}
\label{fig:projected_0}%
\end{figure}

The line-of-sight confusion can be even more severe and some of the
identified filaments can be the result of a chance alignment of
unrelated (or related but still physically separate) density
structures. Of the four filaments selected from the $x$ direction
maps, none is seen as a clear isolated filament when examined from the
other two orthogonal directions. Figure~\ref{fig:projected_0}
illustrates this in the case of filament 1. The filament was
identified from the image towards the direction $x$.
We determined the locations of the volume density maxima for a
few positions along the filament along the line of sight.
The images calculated for the other two orthogonal viewing directions
show that the filament is actually a superposition of at least two
independent density structures. This may be a common scenario also in
the case of real clouds as suggested, for example, by the
identification of several kinematically separate components in some of
the Taurus filaments (Tafalla 2010, private communication).

\begin{figure}
\centering
\includegraphics[width=8.7cm]{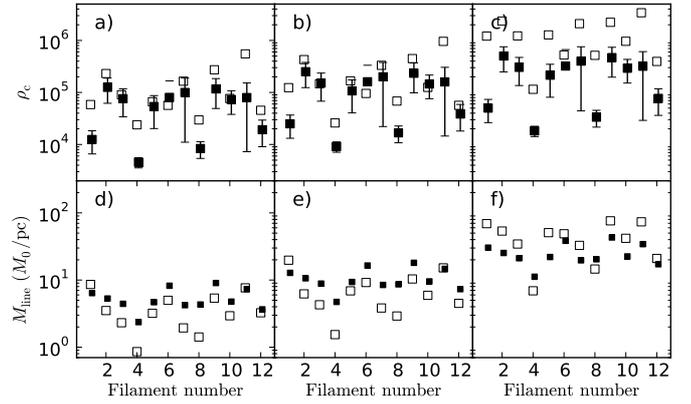}
\caption{
Comparison of the $\rho_{\rm c}$ and filament mass estimates from the
Plummer fits (open symbols) and the actual values of maximum
line-of-sight densities and masses in the cloud model (solid squares).
The error bars on the true densities correspond to the 1 $\sigma$
dispersion of the peak density values along the filament. In the mass
estimates the solid squares correspond to the total mass within a
projected distance of 0.1\,pc of the filament. Model A: frames $a$
and $d$, model B: frames $b$ and $e$, model C: frames $c$ and $f$.
}
\label{fig:compare_true_rho}%
\end{figure}

Figure~\ref{fig:compare_true_rho} shows the central densities of
filaments as derived from the observations. For comparison, the figure
includes the actual maximum line-of-sight density that is obtained
from the model clouds and is averaged along the length of the
projected filament. For model C the densities are overestimated
because of the difference between the true and the assumed
$\kappa_{\nu}$ values. In models A and B there is less bias but
even in model B the densities are sometimes overestimated by a
factor of a few. This is the case for example for the filament number
1 that was already seen to consist of separate density structures
along the line of sight. The result reminds us that continuum
observations cannot directly measure the volume density. The
uncertainty on the $\rho_{\rm c}$ values is several tens of per cent.

Figure~\ref{fig:compare_true_rho} also compares the mass estimates
$M_{\rm line}$ with the actual cloud mass within a projected distance
of 0.1\,pc of the filament centre. Apart from model C, the true
masses are on average slightly higher than the values derived from the
fits of the filament profiles. To some extent this may reflect the
presence of an extended cloud component that in the Plummer fits is
modelled as a separate component and thus is not included in the
$M_{\rm line}$ estimates. In some cases there is clear anticorrelation
between the $\rho_{\rm c}$ and $M_{\rm line}$ values (e.g., filaments 4
and 8). When the fitted profile has a narrow peak, a larger fraction
of the mass gets attributed to the background component. However, the
background is usually low (see Fig.~\ref{fig:A_fwhm3}) and cannot
always explain the factor of $\sim$2 differences in the models A and
B.

The bias in the mass estimates can be traced back to the assumptions
of the column density calculations, i.e., a constant line-of-sight
temperature and a dust spectral index $\beta=2.0$ (instead of the
slightly larger actual value of the dust model). Although the bias in
the column density values should increase towards the centre of the
filament, this was not reflected in the parameters $\rho_{\rm c}$ or
$p$ as a systematic error. In model C the inclusion of dust with
higher sub-millimetre opacity causes bias mainly in the mass estimates
but also the central density and the parameter $p$ should be affected.
The fact that the column densities were overestimated more towards the
centre of the filaments should lead to overestimation of the parameter
$\rho_{\rm c}$. This is also seen in the results, the central
densities being overestimated by a factor of 2--3. This is consistent
with the modified dust being responsible for most of the signal along
the central line of sight. The relative abundance of the dust with
higher $\kappa_{\nu}$ was increased as a function of density and this
could have been expected to lead to higher $p$ values. In
practice, no clear bias is seen.
Together with the high column density, the high dust emissivity
increased the signal-to-noise ratio in model C far above that of
models A and B. Nevertheless, in model C also the scatter of the
parameter values was the largest.

\section{Conclusions}  \label{sect:conclusions}

We have examined the reliability of the parameters that can be derived
for interstellar cloud filaments using dust emission data. The study
is based on synthetic observations calculated with MHD simulations of
filamentary clouds and with radiative transfer modelling of the
sub-millimetre dust emission. The observational parameters
correspond to typical \emph{Herschel} observations. We have examined
the estimated FWHM widths of the filaments and the parameters of the
Plummer functions used to describe the filament profiles. The results
have led to the following conclusions:
\begin{itemize}
\item
For clouds at a distance of $\sim$100\,pc, the profile parameters
obtained from the observations and from the true column density are
usually similar to within a few percent. However, differences up to
tens of percent are observed in the case of a few filaments.
\item 
The results are affected by the instrumental noise. In the higher
density models the signal-to-noise ratios were higher and the errors
were correspondingly smaller. If the instrumental noise
is increased from the assumed default values, the errors increase
rapidly. When the noise is increased by a factor of four, the filament
parameters are no longer well constrained even for nearby clouds.
\item
The uncertainties caused by the background confusion and the 
typical instrumental noise outweigh the radiative transfer effects.
\item
The uncertainties increase rapidly when the resolution of the
observations is decreased. At a distance of $\sim$400\,pc the
filaments are barely resolved, the typical errors are larger by a
factor of a few, and there are more cases where the profile fit leads
to completely wrong parameter values.
\item
We examined a model where the dust opacity is increased as a function
of density. In the analysis the mass estimates are naturally affected
by wrong assumptions of the dust opacity. The errors are larger also
for the other parameters but they are not strongly biased.
\item
The derived filament properties are affected by the line-of-sight
confusion. In our study some of the filaments that were identified
from the column density maps are not continuous structures of the
three dimensional density field. 
\end{itemize}

\begin{acknowledgements}
The authors acknowledge the support of the Academy of Finland grants
No. 127015 and 250741. JM also acknowledges a grant from Magnus Ehrnrooth Foundation.
\end{acknowledgements}

\bibliography{biblio_v2.0}

\end{document}